\documentclass[12pt,a4paper,final]{iopart}

\usepackage{iopams}
\usepackage{graphicx}
\usepackage[breaklinks=true,colorlinks=true,linkcolor=blue,urlcolor=blue,citecolor=blue]{hyperref}
\usepackage[font=footnotesize,labelfont=bf,
   justification=justified,
   format=plain]{caption}

\begin{document}

\title[]{Experimental and Computational Insights Into the Magnetic Anisotropy and Magnetic Behaviour of Layered Room-Temperature Ferromagnet Cr$_{1.38}$Te$_2$}

\author{Shubham Purwar$^{1}$, Tushar Kanti Bhowmik$^1$, Tijare Mandar Rajesh$^2$, Anupam Gorai$^1$, Bheema Lingam Chittari$^2$, and Setti Thirupathaiah$^{1,*}$}
\address{$^1$Condensed Matter and Materials Physics Department, S. N. Bose National Centre for Basic Sciences, Kolkata, West Bengal-700106, India}
\address{$^2$Department of Physical Sciences, Indian Institute of Science Education and Research Kolkata, Mohanpur 741246, West Bengal, India}
\eads{$^{*}$\mailto{setti@bose.res.in}}

\begin{indented}
\item[]\today
\end{indented}

\begin{abstract}
We investigate the structural, magnetocrystalline anisotropy, critical behaviour, and magnetocaloric effect in the layered room-temperature monoclinic ferromagnet Cr$_{1.38}$Te$_2$. The critical behavior is studied by employing various techniques such as the modified Arrott plot (MAP), the Kouvel-Fisher method (KF), and the critical isothermal analysis (CI) around the Curie temperature ($T_C$) of 316 K. The derived critical exponents are self-consistent and obey the rescaling analysis.   The Monte-Carlo simulations reproduce the experimentally obtained critical exponents. However, the derived critical exponents do not suggest any single universality class of the magnetic interactions. On the other hand, the renormalization group (RG) theory suggests 3D-Ising type long-range exchange interactions [$J(r)$], decaying with distance ($r$) as $J (r) = r^{-(d+\sigma)}= r^{-4.73}$. Further, magnetocrystalline anisotropy energy density (K$_u$) is found to be temperature dependent. The ground state magnetic easy-axis ($b$-axis) is identified by analyzing the magnetocrystalline anisotropy energy (MAE) using the density functional theory calculations.  Maximum entropy change -$\Delta S_{m}^{max}$$\approx$2.51 J/kg-K is found near the $T_C$.
\end{abstract}

 \vspace{2pc}

\section{\label{sec:level1}Introduction}

Two-dimensional (2D) van der Waals (vdW) magnetic materials with intrinsic long-range ferromagnetic (FM) ordering have recently received immense research interests due to their potential technological applications in the low-power spintronic devices~\cite{Heron2011,Geim2013,Gong2017,Fei2018,Wang2019,Gibertini2019}. However, their technological applications are limited as the ferromagnetic long-range ordering is destabilized by the dominating thermal fluctuations at higher temperatures~\cite{Mermin1966}. In this regard, strong magnetic anisotropy is required to stabilize the long-range magnetic ordering at room temperature to overcome the thermal fluctuations. Recently, several 2D magnetic materials such as CrI$_3$ (T$_C\approx$ 45 K)~\cite{Huang2017}, Cr$_2$Ge$_2$Te$_6$ (T$_C\approx$ 61 K)~\cite{Liu2017}, and Cr$_2$Si$_2$Te$_6$ (T$_C\approx$ 32 K)~\cite{Williams2015} were synthesized with weak inter-layer coupling leading to strong magnetic anisotropy, but the T$_C$ of these systems is far below the room temperature (RT). Therefore, investigating the layered room-temperature 2D ferromagnetic materials with large magnetocrystalline anisotropy is crucial for potential applications in spintronics and data storage devices~\cite{Zhuang2016,Soumyanarayanan2016,Liu2022}.

On the other hand, theoretically, it was proposed that the layered binary chromium-based telluride, Cr$_{1+x}$Te$_2$ (0$<$x$<$1), are potential candidates to realize the much-anticipated room-temperature 2D ferromagnetism in the bulk~\cite{Zhu2018}. Since then, many Cr$_{1+x}$Te$_2$ based compounds were grown experimentally to study their peculiar 2D magnetism. Generally, Cr$_{1+x}$Te$_2$ possess alternative stacking of CrTe$_2$ (Cr-full) layers intercalated by the Cr-layers (Cr-vacant) within the van der Waals gap between the two CrTe$_2$ layers stacked along the crystal growth axis~\cite{Dijkstra_1989}. Importantly, the intercalated Cr atoms play a vital role in exhibiting various structural, magnetic, and magnetotransport properties depending on the amount of intercalated Cr present in the system. For instance, Cr$_{1+x}$Te$_2$ with x $>$ 0.8 forms into the hexagonal NiAs-type structure~\cite{Ipser1983}, Cr$_{1.33}$Te$_2$ (Cr$_2$Te$_3$) and Cr$_{1.5}$Te$_2$  (Cr$_3$Te$_4$) crystallize into trigonal or monoclinic symmetries~\cite{Hamasaki1975,Dijkstra_1989}, Cr$_{1.25}$Te$_2$  (Cr$_5$Te$_8$) forms into the trigonal phase at higher temperatures and forms into monoclinic phase at low temperatures~\cite{Ipser1983,Chattopadhyay1994}. Interestingly, several recent reports on Cr$_{1+x}$Te$_2$ systems reveal exotic physical properties such as the noncollinear spin-textures~\cite{Wang2019a}, anomalous and topological Hall effects (AHE)~\cite{Yan2019, Tang2022, Liu2022a, Huang2021}, and skyrmion lattice~\cite{Liu2022a, Zhang2022a}. In addition, as the FM vdW materials are best suited for magnetic refrigeration due to their large magnetocaloric effect (MCE),  these systems have also attracted a great deal of research interest from this direction as well~\cite{MorenoRamirez2021}.

In this contribution, magnetic properties, critical behaviour analysis, and magnetocaloric effect have been studied in detail on the high-quality single crystals of monoclinic Cr$_{1.38}$Te$_2$. We systematically investigated the structural properties of  Cr$_{1.38}$Te$_2$ by performing temperature dependent X-ray diffraction measurements and established a relation between the magnetic transitions and the crystal lattice.  The critical exponents were investigated by the modified Arrott plot, the Kouvel-Fisher method, and the critical isotherm analysis. The critical exponents are tested to be self-consistent and obey the rescaling mechanism around the $T_C$. The Monte-Carlo simulations were conducted to reproduce the experimentally obtained critical exponents. Since the critical exponents do not fall into any particular universality class of the magnetic interactions, we employed the renormalization group (RG) theory. The RG theory suggests 3D-Ising type long-range exchange interactions [$J(r)$] in Cr$_{1.38}$Te$_2$, decaying with distance ($r$) as $J (r) = r^{-(d+\sigma)}= r^{-4.73}$. Further, magnetocrystalline anisotropy energy density (K$_u$) is found to be temperature dependent, reaching the maximum (180 k-J/$m^3$) at 110 K. The magnetic easy-axis is identified theoretically by analyzing the magnetocrystalline anisotropy energy (MAE) using the density functional theory calculations.  Further, we have explored the isothermal magnetic entropy change -$\Delta S_{m}$ as a function of the field.

  \begin{figure}[t]
    \centering
    \includegraphics[width=0.8\linewidth]{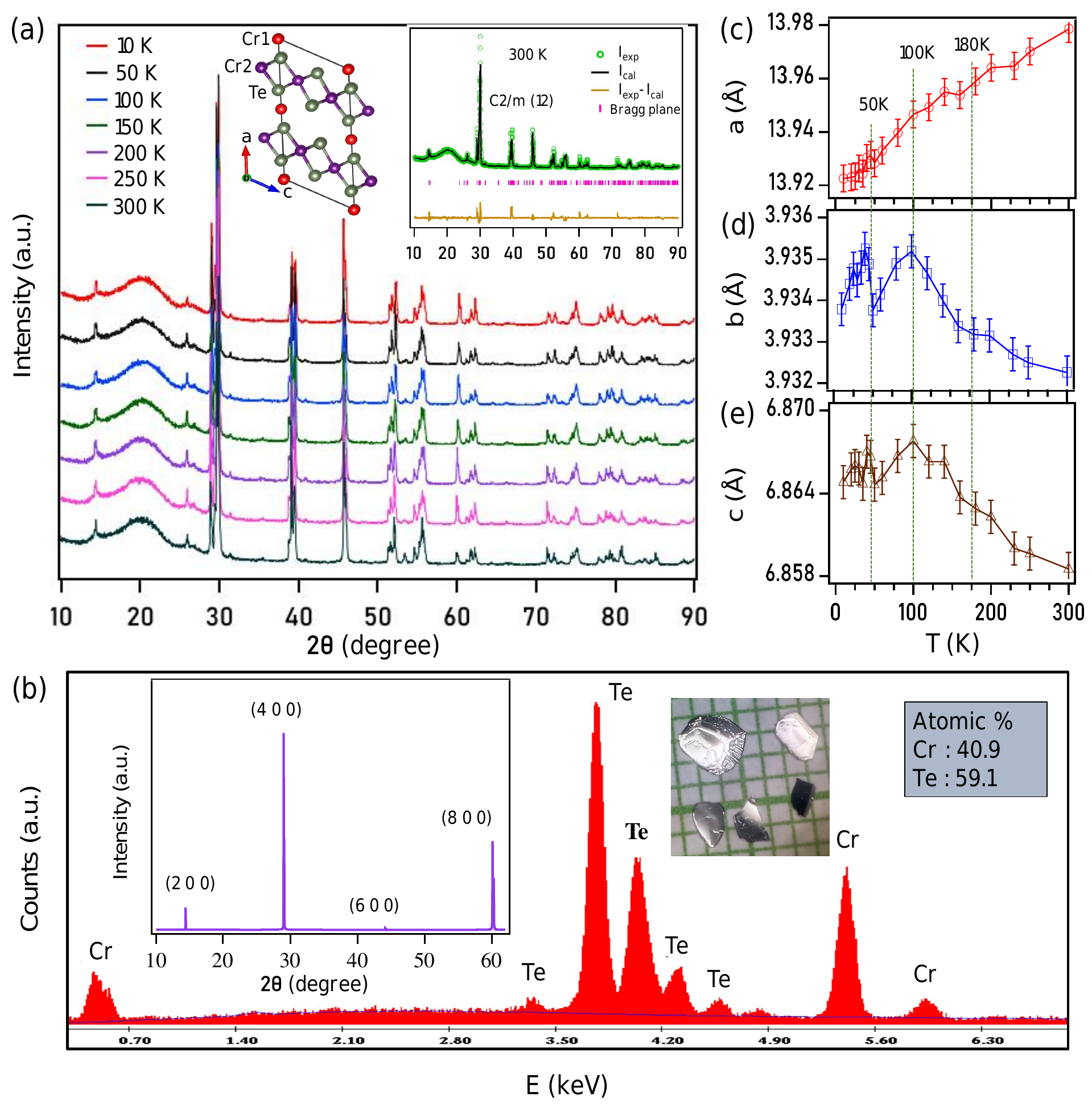}
    \caption{Structural analysis of Cr$_{1.38}$Te$_2$. (a) X-ray diffraction (XRD) measured at different sample temperatures in the range of 10-300 K. Top-right inset in (a) shows XRD pattern measured at T=300 K overlapped with Rietveld refinement. The middle inset in (a) shows a schematic diagram of the crystal structure derived from the Rietveld refinement. (b) Energy dispersive X-ray spectroscopy data showing the actual chemical composition by the atomic percentage. Insets in (b) present X-ray diffraction pattern and photographic image of Cr$_{1.38}$Te$_2$ single crystal. (c), (d), and (e) show temperature dependent lattice parameters $\it{a}$, $\it{b}$, and $\it{c}$, respectively.}
    \label{1}
\end{figure}

\section{Experimental and Computational Methods}

High-quality single crystals were grown using the chemical vapour transport (CVT) method with iodine as a transport agent~\cite{Hashimoto1971}. The exact chemical composition of the as-grown single crystals was determined to be Cr$_{1.38(2)}$Te$_2$ using the energy dispersive X-ray spectroscopy (EDS), indicates a 38\% of excess Cr compared to the nominal composition of CrTe$_2$.  The as-grown single crystals were large in size (4$\times$4 mm$^2$) and were looking shiny. Powder X-ray diffraction (XRD) measurements were performed on the crushed single crystals using Rigaku X-ray diffractometer (SmartLab, 9kW) with Cu K$_\alpha$ radiation of wavelength 1.5406 $\AA$ at various sample temperatures.  Magnetic measurements [$M(T)$ and $M(H)$] were carried out using the physical property  measurement system (9 Tesla-PPMS, DynaCool, Quantum Design).

We performed density functional theory (DFT) calculations to determine the ground state magnetic easy axis in Cr$_{1.38}$Te$_2$ crystal. For this purpose, we used the crystal structure determined from our experimental data. We employed plane-wave basis set as implemented in QUANTUM-ESPRESSO (QE) simulation package~\cite{Giannozzi2009}, with fully relativistic pseudopotentials of Perdew-Burke-Ernzerhof (PBE) type exchange and correlations~\cite{Perdew1996}. We utilized a k-point mesh of 3$\times$8$\times$6 with electronic relaxation convergence upto 1E-9Ry without any external constraints. Spin-orbit coupling (SOC) was included for all the magnetocrystalline anisotropy energy (MAE) calculations. We followed the force theorem as implemented in QE and a two-step process to describe the MAE~\cite{Li2014}. Initially, the self-consistent field (SCF) calculations were performed with scalar-relativistic pseudopotentials to obtain the charge density and the spin moment distribution in real space. After that the spin moment is globally rotated to a desired direction and the spin moment-dependent energy was calculated through non-SCF calculations. The energy difference between two spin moment directions gives the MAE.

Monte-Carlo simulations using the single-flip Metropolis algorithm was conducted to investigate the critical behavior of Cr$_{1.38}$Te$_2$~\cite{Metropolis1953,Binder1993,Newman1999}. To capture the characteristic behaviour, we adapted a discrete 3D Ising-like spin model with the Hamiltonian,

\begin{equation}
H = -\sum_{<i,j>}{J_{ij}S_iS_j} - h\sum_{i}{S_i}
\label{Eq0}
\end{equation}

where, $h$ is the external applied field,  $J_{ij}$ is the nearest neighbour (NN) exchange interaction. $S_i$ and $S_j$ are the total spin at the $i^{th}$ and $j^{th}$ site, respectively. The value of $J_{ij}$ is calculated from the density functional theory (DFT) as discussed in the Supplementary information.

\section{Results and Discussion}

\subsection{Structural Properties}

Figure~\ref{1}(a) shows powder XRD patterns of the crushed Cr$_{1.38}$Te$_2$ single crystals measured at various sample temperatures. The top-right inset in Fig.~\ref{1}(a) shows the Rietveld refinement of the room temperature XRD pattern performed using the Fullprof software~\cite{Rodriguez1993}, confirming the monoclinic crystal structure with a space group of  $C2/m$ (12) without any impurity phases. At room temperature, the lattice parameters were found to be $\it{a}$ = 13.9655(2) \AA, $\it{b}$ = 3.9354(4) \AA, $\it{c}$ = 6.8651(7) \AA, $\alpha = \beta$ = 90$^{\circ}$, and $\gamma$ = 118.326(7)$^{\circ}$. The layered crystal structure of Cr$_{1.38}$Te$_2$ is schematically shown in the middle inset of Fig.~\ref{1}(a). There are two types of Cr [Cr(1) and Cr(2)] atoms present in the unit cell in such a way that the intercalated Cr(1) atoms are sandwiched between two Cr(2)Te$_2$ layers. Fig.~\ref{1}(b) shows EDS spectra of as-grown single crystals, confirming the actual chemical composition of Cr$_{1.38}$Te$_2$ (see the table at the top-right inset). The middle inset in Fig.~\ref{1}(b) shows photographic image of several Cr$_{1.38}$Te$_2$ single crystals. The top-left inset in Fig.~\ref{1}(b) depicts the XRD pattern taken on a Cr$_{1.38}$Te$_2$ single crystal, showing the intensity of (2 0 0) Bragg's plane, confirming that the crystal growth is along the $a$-axis.  Figs.~\ref{1}(c), ~\ref{1}(d), and ~\ref{1}(e) depict the temperature dependent lattice parameters of $\it{a}$, $\it{b}$, and $\it{c}$, respectively. Although we do not observe any low-temperature structural phase transition in Cr$_{1.38}$Te$_2$ as no splitting in the Bragg's peaks is noticed down to 10 K. However, we could observe a monotonic decrease in the lattice constant $\it{a}$ with decreasing temperature. On the other hand, we observe unusual behavior of the lattice constants $\it{b}$ and $\it{c}$ with temperature. Specifically, down to 100 K, both $\it{b}$ and $\it{c}$ monotonically increase with decreasing temperature. But below 100 K, $\it{b}$ and $\it{c}$ decrease with decreasing temperature to have a minimum at 50 K and again increase with decreasing temperature to peak out at 35 K. Worth to mention here that at around 180$\pm$20 K we find a noticeable change in the lattice constants $\it{a}$, $\it{b}$ and $\it{c}$.    Such an unusual lattice constants behavior with temperature plausibly due to the magneto-volume effect in these type of systems Cr$_{1+x}$Te$_2$~\cite{Kubota2023,Li2022}.

\begin{figure}[t]
    \centering
    \includegraphics[width=0.6\linewidth]{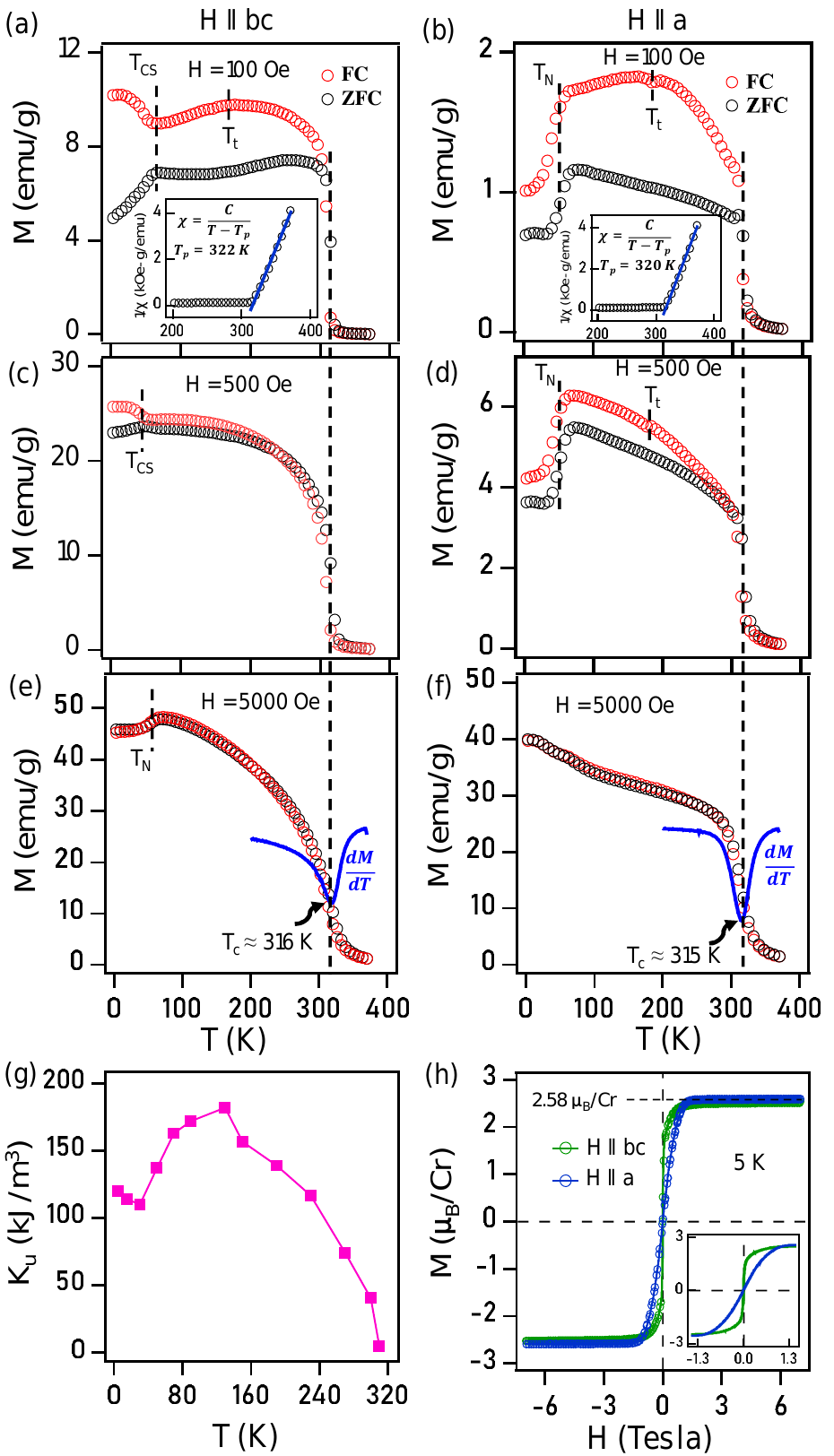}
    \caption{Temperature dependent magnetization $M(T)$ measured under zero-field-cooling (ZFC) and field-cooling (FC) modes at different applied magnetic fields ($H$). (a)-(f) show the $M(T)$ measured at various fields for $H\parallel bc$ and $H\parallel a$. Insets in (a) and (b) show the fittings with Curie-Weiss law. Here $T_C$ is the Curie temperature, $T_N$ is the N$\acute{e}$el temperature, $T_{CS}$ is the spin-canting related transition, and $T_t$ is lattice related transition.  Blue curves in (e) and (f) represent $dM/dT$ in which the peak-dip corresponds to the Curie temperature ($T_C$). (g) Magnetocrystalline anisotropy energy density ($K_u$) is plotted as a function of temperature. (h) Magnetization isotherms $M(H)$ measured at T = 5 K for $H\parallel bc$ and $H\parallel a$. Inset in (h) shows the zoomed-in image demonstrating the magnetic anisotropy between $H\parallel bc$ and $H\parallel a$.}
    \label{2}
\end{figure}

\subsection{Magnetic Properties}

Magnetization of Cr$_{1.38}$Te$_2$ plotted as a function of temperature [$M(T)$], measured at a magnetic field ($H$) of 100 Oe, applied perpendicular ($H\parallel bc$) and parallel ($H\parallel a$) to the $a$-axis is shown in Figs.~\ref{2}(a) and ~\ref{2}(b), respectively taken in the field-cooled (FC) and zero-field-cooled (ZFC) modes. Figs.~\ref{2}(c)-(d) and Figs.~\ref{2}(e)-(f) are same as Figs.~\ref{2}(a)-(b) except that the former data are measured at $H=500$ Oe and the latter data are measured at $H=5000$ Oe. From Fig.~\ref{2}, we observe a high-temperature paramagnetic (PM) to a low-temperature ferromagnetic (FM) transition at around a Curie temperature of $T_C\approx315\pm0.5$ K,  which is close to the previously reported value of 316 K on a similar system~\cite{Yamaguchi1972}. In the ferromagnetic state, we observe a significant thermal irreversibility between ZFC and FC modes for the low applied fields [see Figs.~\ref{2}(a) and \ref{2}(b)], indicating the presence of magnetocrystalline anisotropy (MCA) in the system~\cite{Liu2019, Kumar2000}. In addition, we notice another magnetic transition at around 50 K for $H\parallel bc$ and $H\parallel a$. While the magnetic transition for  $H\parallel bc$ looks more like a spin-canting effect, the magnetic transition for $H\parallel a$ is an antiferromagnetic type. Such complex low-temperature directional-dependent magnetic transitions can be attributed to the competition between FM and canted-AFM phases~\cite{Bertaut1964,aZhang2020, Huang2008}. Further, with increasing applied magnetic field, the spin-canting type in-plane ($bc$-plane) magnetic transition gradually dissipates [see Figs.~\ref{2}(b) and ~\ref{2}(c)] and becomes a clean AFM-type at higher applied fields. Similarly, the out-of-plane ($a$-axis) AFM transition gradually dissipates with increasing applied field and completely disappears at $H=5000$ Oe. In support of the magneto-volume effect in these systems~\cite{Kubota2023,Li2022}, we could clearly find a change in magnetization $M(T)$ (FC data) at around $T_t= 180$ K where we also observe a noticeable change in the lattice constants [see Figs.~\ref{1}(c)-(d)].

\begin{figure*}[t]
    \centering
    \includegraphics[width=\linewidth]{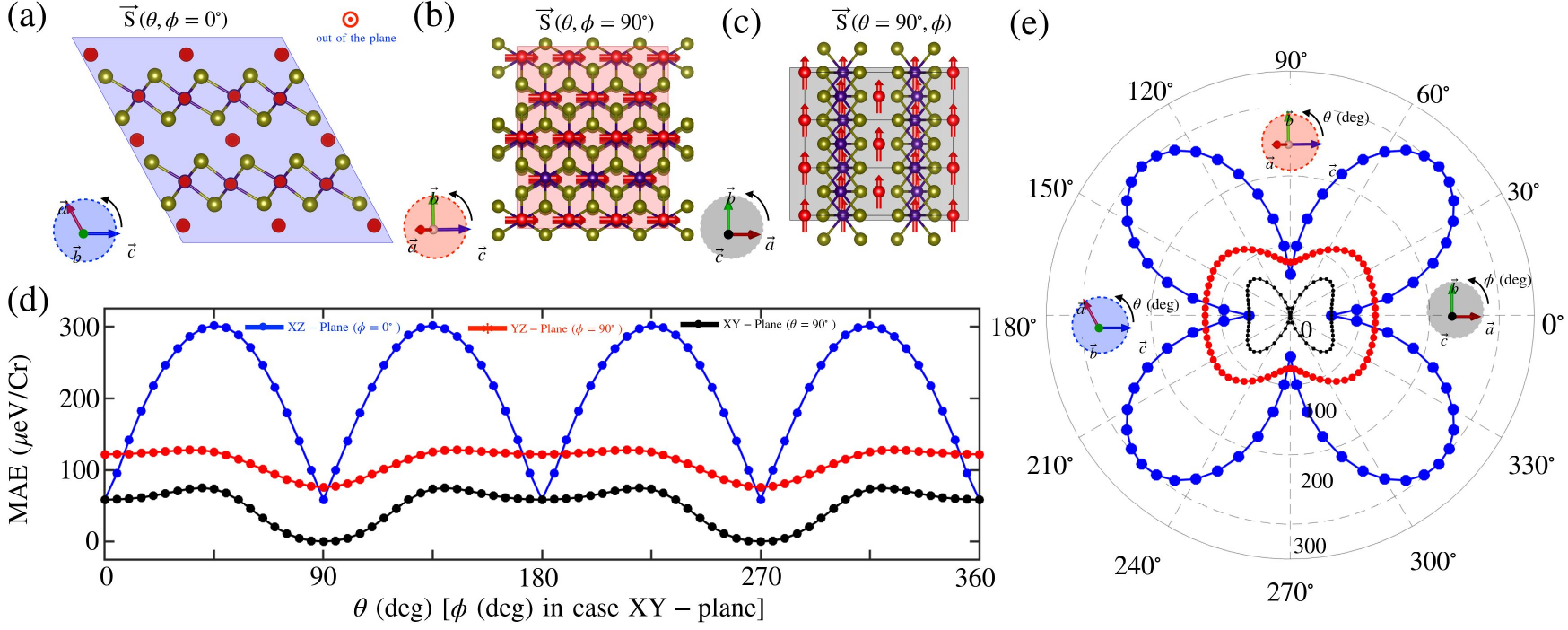}
    \caption{The magnetocrystalline anisotropy energy (MAE) of Cr$_{1.38}$Te$_2$: The spin direction of Cr atoms, denoted as S($\theta$), is defined as $\theta$ = 0$^\circ$ along the Z-axis. The inclined spin direction ($\theta$ $\neq$ 0) projects onto the XY-plane, with the projection measured from the X-axis at an angle ($\phi$).  $\phi$ = 0$^\circ$ aligns with the X-axis. For $\theta$ $\neq$ 0, the spin moment direction with $\phi$ = 0$^\circ$ is confined to the XZ-plane, and for $\theta$ $\neq$ 0 and $\phi$ = 90$^\circ$, the spin rotation is restricted to the YZ-plane. XY-plane spin rotation occurs at $\theta$ = 90$^\circ$ with $\phi$ $\neq$ 0. The spin rotation (S) of Cr atoms in Cr$_{1.38}$Te$_2$ is described in three planes (a) XZ (ac)-plane, (b) YZ (bc)-plane, and (c) XY (ab)-plane. (d) MAE per Cr atom for Cr$_{1.38}$Te$_2$ concerning spin rotation in three different planes. (e) MAE plotted in polar graph for a better visualization.}
    \label{3}
\end{figure*}

Previous neutron diffraction studies on Cr$_{1.5}$Te$_2$ suggested a mixed spin-state  of  Cr$^{2+}$ (Cr1) and Cr$^{3+}$ (Cr2), leading to a canted spin structure due to the ferromagnetic components of Cr1 and Cr2 tilted by 30$^o$ and 38$^o$, respectively, about the $\it{a}$-axis and the antiferromagnetic (AF) components of Cr1 and Cr2 are aligned in the $\it{bc}$-plane~\cite{ANDRESEN1970}. Therefore, the competition between FM and AFM phases could be playing a vital role in garnering the multiple magnetic transitions, in addition to the canted spin structure, in these systems~\cite{Zhang2022, Chen2023}. The inverse susceptibility ($1/\chi$) plotted as a function of temperature, as shown in the inset of Figs.~\ref{2}(a) and ~\ref{2}(b),   for both field directions follow the Curie-Weiss law $\chi(T)=C/(T-T_p)$. Here $\chi$ is magnetic susceptibility, $C$ is the Curie constant, and $T_p$ is the Curie-Weiss temperature. From the fittings we obtained T$_p$ that are greater than $T_C$ for both directions, indicating a predominant FM exchange interactions in this system. We further calculated the effective magnetic moment in the paramagnetic region using the relation $\mu_{eff} \approx 2.84\sqrt{MC}$~\cite{Makovetskii1978}, where M is a molar-mass (in gram). The calculated effective moment ($\mu_{eff}$) is 4.531 $\mu_B$ for $H\parallel a$ and 4.403 $\mu_B$ for $H\parallel bc$. The values of  $\mu_{eff}$ lies between the theoretical predictions of 3.87 $\mu_B$  for Cr$^{3+}$ and 4.90 $\mu_B$ for Cr$^{2+}$, the mixed valence state existing in this system is consistent with previous reports~\cite{Bertaut1964,ANDRESEN1970,Shimada1996}.

\begin{figure*}[t]
    \centering
    \includegraphics[width=\linewidth]{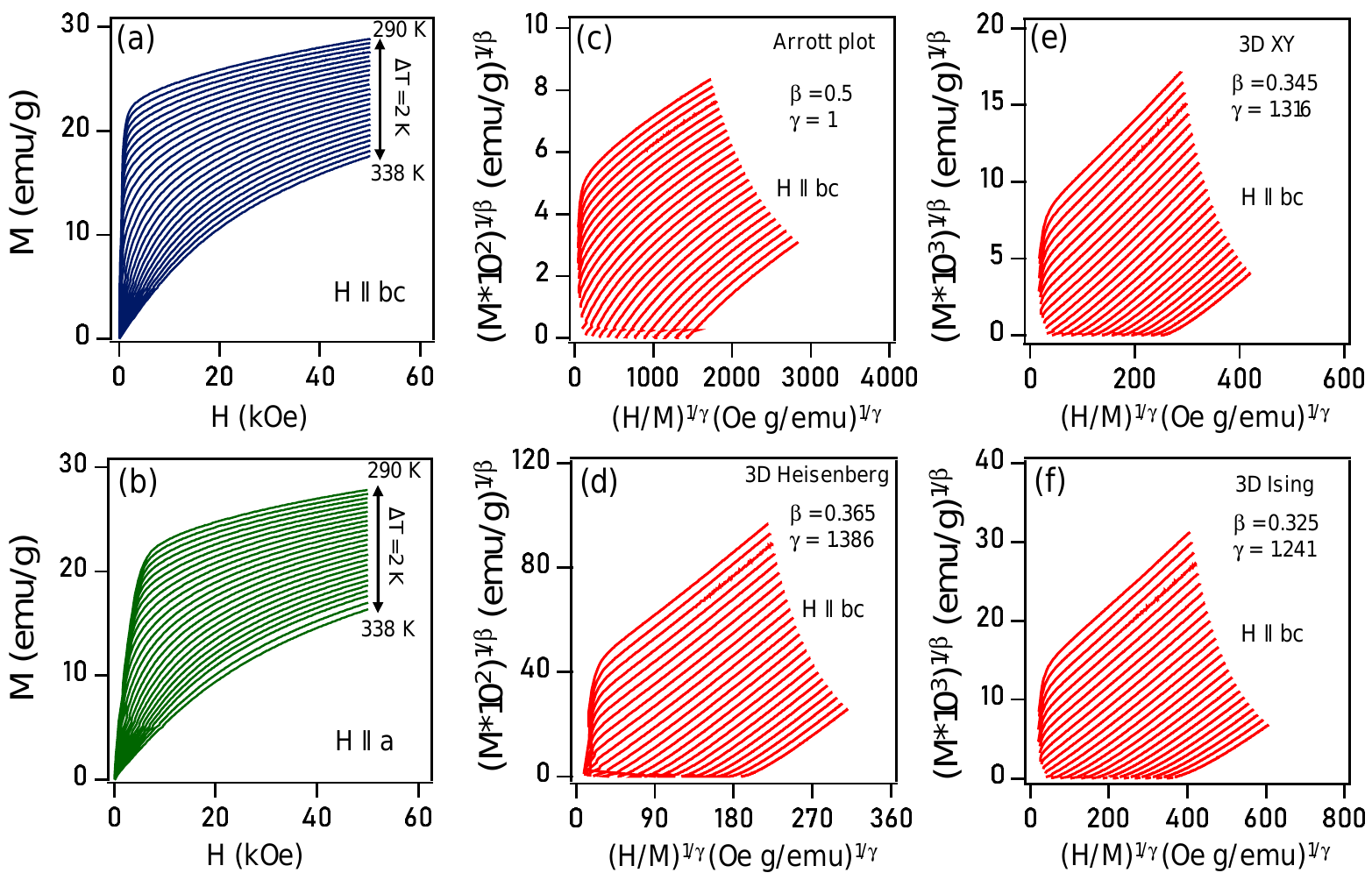}
    \caption{Isothermal magnetization $M(H)$ measured between 290 and 338 K at an interval of $\Delta T=2$ K for (a) $H\parallel bc$ and (b) $H\parallel a$. (c) Arrott plots of M$^2$ vs H/M. Modified Arrott plot (Arrott-Noaks plot) of M$^{1/\beta}$ vs (H/M)$^{1/\gamma}$ for (d) 3D Heisenberg model, (e) 3D XY model, and (f) 3D Ising model done for the easy-plane of magnetization ($bc$-plane).}
    \label{4}
\end{figure*}

Further, the magnetocrystalline anisotropy energy density $K_u$ can be obtained using the Eqn.~\ref{Eq13}. Fig.~\ref{2}(g) depicts $K_u$ plotted as a function of temperature. We can observe that $K_u$ monotonically increases with decreasing temperature from 320 K down to 110 K. But below 110 K,  due to the additional magnetic transition as noticed in Fig.~\ref{2}, $K_u$ decreases with decreasing temperature. Usually during the disorder to order magnetic transition (PM-FM), the Cr$_{1+x}$Te$_2$-based systems exhibit a negative volume thermal-expansion leading to strong magnetoelastic coupling~\cite{Xu2020}. The same could be true in our studied system of Cr$_{1.38}$Te$_2$ as well, as we observe fluctuating lattice constants below 110 K [see Figs.~\ref{1}(c-e)]. Nevertheless, the obtained K$_u$ = 120 kJ/m$^3$ at 2 K in Cr$_{1.38}$Te$_2$ is much higher than the K$_u$ values of some of the other 2D magnetic systems such as CrBr$_3$ ($\approx$ 86 kJ/m$^3$ at 5 K)~\cite{Richter2018}, Cr$_2$Ge$_2$Te$_6$ ($\approx$ 20 kJ/m$^3$ at 2 K)~\cite{Liu2019a}, and Cr$_2$Si$_2$Te$_6$ ($\approx$ 65 kJ/m$^3$ at 2 K)~\cite{Liu2019a}. Further, the large $K_u\approx 180$ kJ/m$^3$ at 110 K with a Curie temperature of 315 K observed in Cr$_{1.38}$Te$_2$ poses potential room-temperature technological applications in magnetic storage devices~\cite{Sbiaa2011}, spin valves~\cite{Cortie2020}, magnetic tunnel junctions~\cite{Cortie2020}, and spin-transfer torque devices~\cite{Krizakova2021}.


\begin{equation}
    K_u = \mu_0\int_{o}^{M_s}[H_{bc}(M)-H_{a}(M)]dM
    \label{Eq13}
\end{equation}

Fig.~\ref{2}(h) depicts the magnetization isotherms $M(H)$ measured for $H\parallel bc$ and $H\parallel a$ at T=5 K, respectively, suggesting the $bc$-plane as the easy-plane of magnetization. Further, since we do not observe significant hysteresis in the $M(H)$ curves, Cr$_{1.38}$Te$_2$ can act as a soft ferromagnet. The in-plane and out-of-plane saturation magnetization (M$_S$) are found to be 2.55 $\mu_B$ and 2.586 $\mu_B$, respectively, which are smaller compared to the free-Cr ion (3.0 $\mu_B$). This discrepancy may be ascribed to the itinerant nature of Cr$_{1.38}$Te$_2$~\cite{Liu2017a}. To check the degree of itinerancy in Cr$_{1.38}$Te$_2$, we employed the itinerant magnetism model by Rhodes and Wohlfarth~\cite{Wohlfarth1978, Moriya1979}. In this method, one estimates the Rhodes-Wohlfarth ratio (RWR), $M_c/M_S$.   Here M$_S$ is the saturation magnetization and M$_c$ is the ground state magnetic moment which is calculated using the equation M$_c$(M$_c$ + 2) = $\mu_{eff}^2$. In this way, we estimated RWR=1.89 for $H\parallel a$ and RWR = 1.82 for $H\parallel bc$. It is known that for the localized magnetic systems $RWR=1$ and for the itinerant systems $RWR>1$~\cite{Wohlfarth1978,Moriya1979}. Since estimated RWR values of Cr$_{1.38}$Te$_2$ are greater than one, it must be an itinerant magnetic system.

To further elucidate the magnetic structure of Cr$_{1.38}$Te$_2$, we performed DFT calculations as shown in Fig.~\ref{3}.  We calculated the magnetocrystalline anisotropy energy (MAE) of Cr$_{1.38}$Te$_2$ for three different spin structures projected onto the $ac$-, $bc$-, and $ab$-planes as described in the Figs.~\ref{3}(a), ~\ref{3}(b), and ~\ref{3}(c), respectively. In Fig.~\ref{3}(d), the blue-, red- and black-curves show the MAE plotted for the spin configurations shown in Figs.~\ref{3}(a), ~\ref{3}(b), and ~\ref{3}(c), respectively. From Fig.~\ref{3}(d), it is clear that the spin configuration shown Fig.~\ref{3}(c) has the minimum MAE at $\theta= \phi=90^{\circ}$. That means, the ground state magentic easy-axis of Cr$_{1.38}$Te$_2$ is parallel to the $b$-axis, which is precisely predicted compared to the experimental observation of $bc$-plane.  It is further interesting to note that the projected spin configurations onto the $ab$- and $ac$-planes have the same magnetocrystalline anisotropy energy along the $c$- and $a$-axes.  Further, maximum energy is found at $\theta=45^{\circ}$ for all the spin configurations and the highest MAE is found for the spin configuration shown in Figs.~\ref{3}(c). Fig.~\ref{3}(e) depicts the angle-dependent MAE plotted in the polar graph for a better visualization. From Fig.~\ref{3}(e) we can notice that the magnetocrystalline anisotropy energy of the spin configuration shown in Fig.~\ref{3}(a) has the four-fold symmetry, while the MAE of the spin configurations shown in Figs.~\ref{3}(b) and \ref{3}(c) have the two-fold symmetry.

\begin{figure*}[t]
    \centering
    \includegraphics[width=\linewidth]{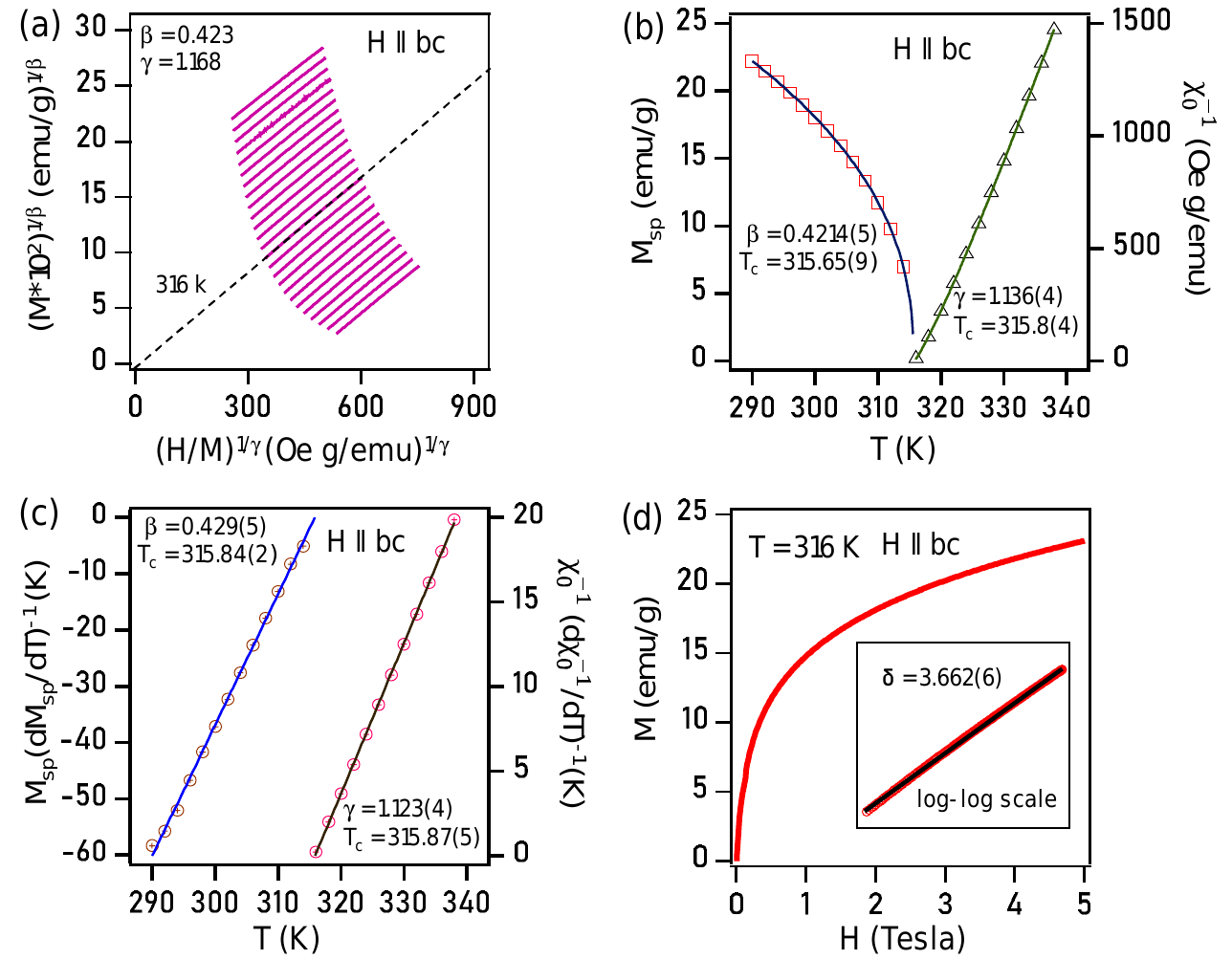}
    \caption{(a) Modified Arrott plot of M$^{1/\beta}$ vs (H/M)$^{1/\gamma}$ done at high field values with critical parameters, $\beta$ = 0.423 and $\gamma$ = 1.168, (b) Temperature dependent spontaneous magnetization M$_{sp}$(T) (left axis) and inverse initial susceptibility $\chi_0^{-1}(T)$ (right axis). Solid curves in (b) are the fits by Eqns.~\ref{3} and \ref{4}. (c) Kouvel-Fisher plots as a function of temperature,  $M_{sp}(T)[dM_{sp}(T)/dT]^{-1}$ (left axis) and $\chi_0^{-1}(T)[d \chi_0^{-1}(T)/dT]^{-1}$ (right axis). Solid lines in (c) are the fits by Eqns.~\ref{6} and \ref{7}, (d) Isothermal magnetization $M(H)$ plot at $T_C$ = 316 K. Inset shows the plot in $log-log$ scale, fitted with the Eqn.~\ref{5}.}
    \label{5}
\end{figure*}

\subsection{Critical Analysis}

To understand the nature and dimensionality of the ferromagnetism in Cr$_{1.38}$Te$_2$, we investigated the critical exponents $\beta$, $\delta$, $\gamma$ which are coupled to the spontaneous magnetization ($M_{SP}$) below $T_C$, inverse initial susceptibility $\chi_0^{-1}(T)$  above $T_C$ , and magnetization isotherms [$M(H)$] at $T_C$, respectively. In this regard, magnetization isotherms $M(H)$ measured within the temperature of 290 to 338 K with a step size of 2 K for both $H\parallel bc$ and $H\parallel a$ are shown in Figs.~\ref{4}(a) and ~\ref{4}(b), respectively. Fig.~\ref{4}(c) shows Arrott plots ($M^2~vs~H/M$) obtained from the $M(H)$ data.  Usually Arrott plots show a set of parallel lines of various temperatures in the high field region~\cite{Arrott1957}. At $T_C$, positive slope of the Arrott plot suggest second-order magnetic transition, while the negative slope of the Arrott plot suggest first-order magnetic transition~\cite{Banerjee1964}. Since the Arrot plot slope at $T_C$ is positive in Fig.~\ref{4}(c), the ferromagnetic transition observed in Cr$_{1.38}$Te$_2$ is a second-order phase transition. On the other hand, the scaling hypothesis suggests that the second-order phase transition around $T_C$ can be described by a set of critical exponents and equations of the magnetic state~\cite{Stanley1971}. Moreover, in the vicinity of second-order phase transition, the divergence of correlation length $\zeta$ = $\zeta_0$[(T - T$_c$)/T$_c$]$^{-\nu}$ leads to the universal scaling laws~\cite{Stanley1971, Fisher1967}.  The mathematical equations involving the critical exponents are given by,

\begin{equation}
M_{sp}(T) = M_0(-\epsilon)^\beta, for \epsilon < 0, T < T_c
\label{Eq1}
\end{equation}

\begin{equation}
\chi_0^{-1}(T) = (h_0/m_0)\epsilon^\gamma, for \epsilon> 0, T > T_c
\label{Eq2}
\end{equation}

\begin{equation}
 M = DH^{1/\delta}, for \epsilon = 0, T = T_c
 \label{Eq3}
\end{equation}

where $\epsilon$ = (T – T$_C$)/T$_C$ is the reduced temperature, $M_0$, $h_0/m_0$, and $D$ are the critical amplitudes~\cite{Fisher1967}. To obtain the critical exponents $\beta$, $\gamma$, and $\delta$ as well as the exact value of $T_C$,  we first tried the modified Arrott plot (MAP) technique and plotted  M$^{1/\beta}$ vs. (H/M)$^{1/\gamma}$ following different models around $T_C$. Fig.~\ref{4}(d) shows MAP for 3D Heisenberg model ($\beta = 0.365, \gamma = 1.386$)~\cite{Kaul1985}, Fig.~\ref{4}(e) shows MAP for 3D XY model ($\beta = 0.345, \gamma = 1.316$)~\cite{LeGuillou1980}, and Fig.~\ref{4}(f) shows MAP for 3D Ising model ($\beta = 0.325, \gamma = 1.241$)~\cite{Kaul1985}. Some other models like 2D-Ising model ($\beta = 0.125, \gamma = 1.75$)~\cite{Fisher1972} and Tricritical mean field model ($\beta = 0.25, \gamma = 1$)~\cite{Kim2002} were also studied (not shown). It can be seen that none of these models can produce the required parallel lines in the high field region to properly explain the magnetic interactions present in Cr$_{1.38}$Te$_2$.

 \begin{figure*}[htbp]
    \centering
    \includegraphics[width=\linewidth]{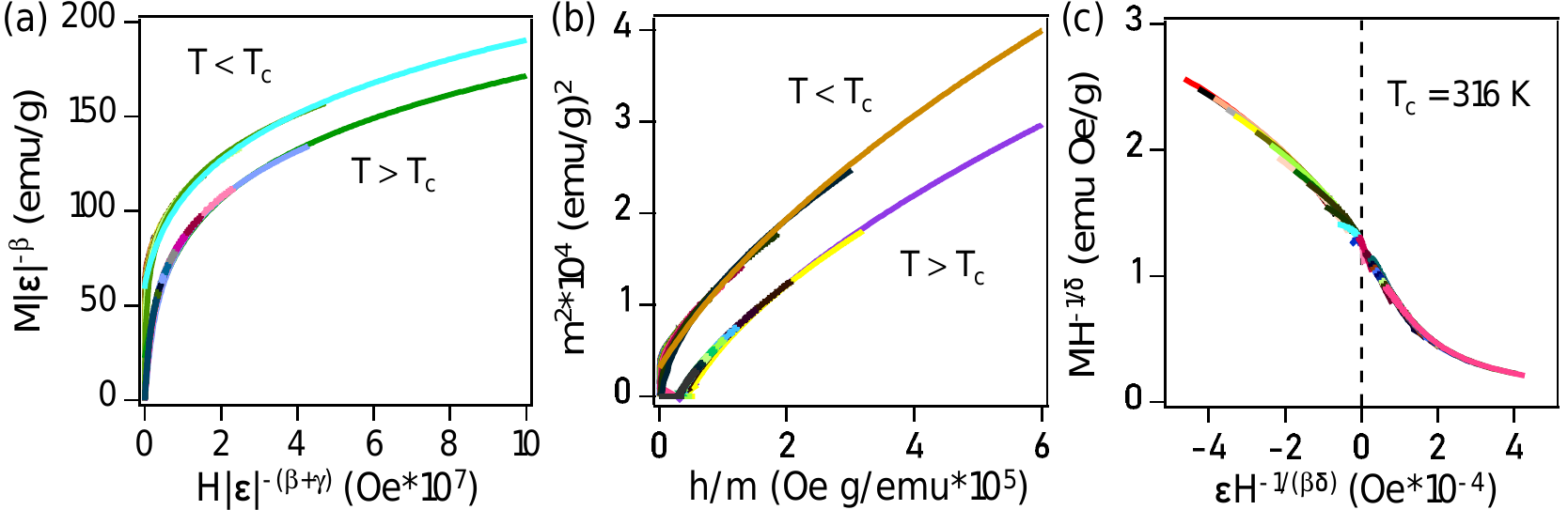}
    \caption{(a) and (b) are the plots of normalized magnetization $'m'$ as a function of re-normalized field $'h'$ below and above $T_C$. (c) Rescaling of $M(H)$ curves by MH$^{-1/\delta}$ vs $\epsilon{H^{-1/\beta\delta}}$.}
    \label{6}
\end{figure*}

Therefore, we used modified Arrott plot (MAP) technique iteratively and plotted M$^{1/\beta}$ vs. (H/M)$^{1/\gamma}$ as shown in Fig.~\ref{5}(a).  This method produces the required parallel lines of all the measured temperatures. Thus, $\chi_0^{-1}(T)$ and M$_{sp}(T)$ are the intercepts on the $x$-axis and $y$-axis of these linear plots, respectively~\cite{Arrott1967,Pramanik2009}.   The critical exponents obtained in this method are given by $\beta$=0.423 and $\gamma$=1.168. Fig.~\ref{5}(b) depicts M$_{sp}$ and $\chi_0^{-1}$ plotted as a function of temperature derived from Fig.~\ref{5}(a). From the Eqns.~\ref{Eq1} and \ref{Eq2} we estimate the critical exponents $\beta$ = 0.421(5) with $T_C$ = 315.65(9) K and $\gamma$ = 1.136(4) with $T_C$ = 315.8(4) K. The estimated $T_C$ values are very close to the value of 315$\pm$0.5 K obtained from the $M(T)$ data [see Fig.~\ref{2}]. Further, the critical exponents can be found more accurately using the Kouvel-Fisher (KF) plots~\cite{Kouvel1964}, involving the equations,

\begin{equation}
M_{sp}(T)[dM_{sp}(T)/dT]^{-1} = (T - T_c)/\beta,
\label{Eq4}
\end{equation}

\begin{equation}
 \chi_0^{-1}(T)[d \chi_0^{-1}(T)/dT]^{-1} = (T - T_c)/\gamma,
 \label{Eq5}
\end{equation}

Fig.~\ref{5}(c) depicts the KF plots fitted with the Eqns.~\ref{Eq4} and \ref{Eq5} below and above $T_C$. The derived critical exponents in the KF method are given by $\beta$ = 0.429(5) with $T_C$ = 315.84(2) K and $\gamma$ = 1.123(4) with $T_C$ = 315.87(5) K, respectively. The values of $\beta$, $\gamma$, and T$_C$ obtained from the KF method are in good agreement with the values obtained from the MAP method, confirming that critical exponents are reliable and intrinsic.

In addition, the exponent $\delta$ has been calculated using the Widom scaling relation, $\delta$ = 1 + $(\gamma$/$\beta)$~\cite{Widom1964}. We estimate $\delta$ = 3.700(6) and $\delta$ = 3.617(8) from the MAP and the KF plots, respectively. Importantly, the $\delta$ values obtained from both MAP and KF methods are close to the value $\delta$ = 3.662(6) obtained from the critical isotherm (CI) method following the Eqn.~\ref{Eq3} as shown in Fig.~\ref{5}(d).  Therefore, the critical exponents derived by different approaches are self-consistent and reasonably accurate within the experimental precision. Reliability of the derived critical exponents and the critical temperature has been examined by the scaling theory. For the magnetic materials,  in the critical asymptotic region, the scaling equation of state is given by~\cite{Stanley1971},
 \begin{equation}
        M (H, \epsilon) = \epsilon^\beta f_\pm (H/\epsilon^{\beta + \gamma})
        \label{Eq6}
 \end{equation}
or
\begin{equation}
    m = f_\pm (h)
     \label{Eq7}
\end{equation}

\begin{table}[htbp]
\caption{Comparison of the critical exponents of Cr$_{1.38}$Te$_2$ with different theoretical models.}
\footnotesize
\begin{tabular}{c @{\extracolsep{\fill}} ccccccc}
 \br
Composition & Technique & Crystal Structure & $\beta$ & $\gamma$ & $\delta$ & Reference \\ [1.5ex]
\mr
Cr$_{1.38}$Te$_2$ & MAP & Monoclinic & 0.421(5) & 1.136(4)& &This work \\ [1.2ex]
&KF method & Monoclinic & 0.429(5) & 1.123(4)& &This work  \\[1.2ex]
&Critical Isotherm & Monoclinic &  & & 3.662(6) & This work \\[1.2ex]
&Monte-Carlo Simulation &  & 0.413(1) & 1.175(2)& &This work  \\[1.2ex]
Theory & Landau mean field  &  & 0.5 & 1 & 3 & \cite{Arrott1957} \\[1.2ex]
 Theory & 3D Heisenberg  &  & 0.365 & 1.386 & 4.80 & \cite{Kaul1985}\\ [1.2ex]
 Theory & 3D Ising  &  & 0.325 & 1.241 & 4.82 & \cite{Kaul1985}\\[1.2ex]
\br
\end{tabular}
\label{T1}
\end{table}
\normalsize

where $m = \epsilon^{-\beta} M (H, \epsilon) $ is normalized magnetization and $h =  H\epsilon^{-(\beta + \gamma)}$ is the normalized field. Now, if the $\beta$ and $\gamma$ values derived from the MAP and KF methods are appropriately chosen, the scaled magnetization $m$ and field $h$ should fall onto two different branches of the universal curves following the Eqns.~\ref{Eq6} and \ref{Eq7}. One branch is for $T<T_C$ and another branch is for $T>T_C$, as shown in Figs.~\ref{6}(a) and \ref{6}(b). This suggests that the magnetic interactions get properly re-normalized in the vicinity of $T_C$. In addition, the scaling equation of state can also be expressed by,

\begin{equation}
    \frac{H}{M^\delta} = k\frac{\epsilon}{H^{1/\beta}}
    \label{Eq9}
\end{equation}

where $k$ is the scaling constant. Thus, all the experimental data collapse into a single curve following the Eqn.~\ref{Eq9} as shown in Fig.~\ref{6}(c). In Fig.~\ref{6}(c), $T_C$ is located at the zero point of the $x$-axis. Therefore, the scaling hypothesis reaffirms the reliability of the derived critical exponents.

The obtained critical exponent $\beta$=0.429(5) is in between the 3D Heisenberg-type ($\beta$=0.365) and the Landau mean-field-type ($\beta$=0.5) magnetic interactions, whereas $\gamma$=1.123(4) is close to the 3D Ising model ($\gamma$=1.241). This indicates that Cr$_{1.38}$Te$_2$ does not fall into any particular universality class of the ferromagnetism, consistent with the complex magnetic interactions reported in previous studies on the similar systems~\cite{Wang2023, Goswami2024}, except for that the reported $\beta$ values are closer to the 3D Heisenberg-type.   On the other hand, the renormalization group (RG) theory suggests that the long-range exchange interactions decay with the distance $r$ as $J (r) = r^{-(d + \sigma)}$ and short-range exchange interactions decay as $J (r) = e^{-r/b}$, respectively~\cite{Fisher1967,Fisher1972}. Here, $d$ is the spatial dimensionality, $\sigma$ is a positive constant, and $b$ is a spatial scaling factor. The relation between the critical exponent $\gamma$ and $\sigma$ can be expressed by,

 \begin{equation}
  \gamma = 1 + \frac{4}{d}\frac{(n + 2)}{(n + 8)}\Delta\sigma +\frac{8(n - 4)(n + 2)}{(n + 8)^2d^2}\\
         * [\frac{2(7n + 20)G(d/2)}{(n + 8)(n - 4)}+1 ]\Delta\sigma^2
   \label{Eq8}
\end{equation}

where $\Delta\sigma = \sigma - d/2$, $ G(d/2) = 3 -0.25*(d/2)^2$. $n$ and $d$ are the spin and spatial dimensionality of the system. RG theory analysis proposes that the spin interaction is either short or long-range depending on $\sigma>$2 or $\sigma<$2, respectively. By considering the experimental $\gamma$ value of 1.123 from the KF method, following the RG theory, we have calculated the range of exchange interaction $\sigma$ and critical exponent $\beta$ following the equations, $\nu=\gamma/\sigma$, $\alpha=2-\nu*d$, $\beta=(2-\alpha-\beta)/2$, where $\nu$ and $\alpha$ are the exponents of correlation length. This is done for various sets of ${d : n}$ values. We find that the critical exponents derived using ${d : n}={3:1}$ matches well with the values obtained from the KF method.  From this, we suggest that Cr$_{1.38}$Te$_2$ is close to the 3D-Ising type exchange interactions with long-range order decaying as $J (r) = r^{-(d + \sigma)}= r^{-4.73}$ for $d=3$ and $\sigma$=1.73.

 \begin{figure*}[ht]
    \centering
    \includegraphics[width=\linewidth]{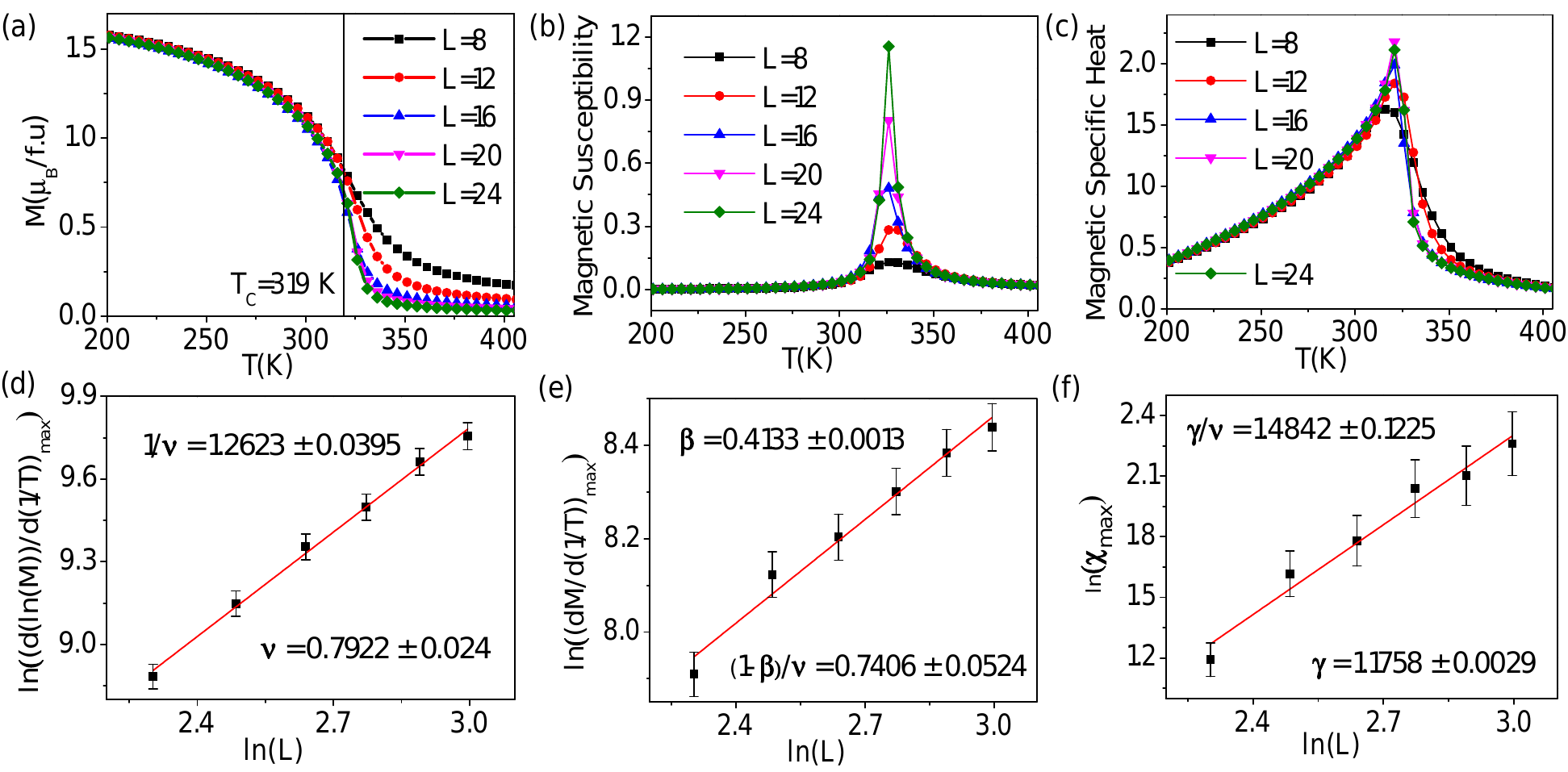}
    \caption{(a) Temperature dependent magnetization for different lattice sizes ($L$), (b) Magnetic susceptibility per spin plotted as a function of temperature, (c) Magnetic specific heat plotted as a function of temperature. $ln-ln$ plot of (d) $[\frac{d(lnM)}{d(1/T)}]_{max}$, (e) $[\frac{dM}{d(1/T)}]_{max}$, (f) $\chi_{max}$. Red-colored line in (d), (e), and (f) are the linear fits.}
    \label{7}
\end{figure*}

In Cr$_{1+x}$Te$_2$ systems, the crystal structure features an alternating arrangement of a fully occupied CrTe$_2$-layer and a Cr-vacant layer along its growth direction. Tuning the $\it{x}$ value not only influences the crystal symmetry,  but also changes the spin configuration of the system~\cite{Liu2010,Liu2017a,Mondal2019, Cao2019,Zhang2020,Fujisawa2020,Fujisawa2023}. Therefore, several studies exist on the critical analysis of Cr$_{1+x}$Te$_2$ systems and found differing exchange interactions for different Cr concentrations and crystal structures. For instance, the trigonal CrTe$_2$, trigonal Cr$_{1.25}$Te$_2$, the monoclinic Cr$_{1.5}$Te$_2$ exhibit 3D Ising-type magnetic interactions~\cite{Liu2017a,Zhang2018,Zheng2023,Wang2023}, bulk and thin-film of trigonal Cr$_{1.6}$Te$_2$ follow 3D Heisenberg and mean-field type interactions, respectively~\cite{Zhang2020,Wang2022}, thin-film hexagonal Cr$_{2}$Te$_2$ displays 2D Heisenberg type interactions~\cite{Liu2023}.   

In order to be sure on the universality class of Cr$_{1.38}$Te$_2$, derived using the RG theory with the help of experimental critical exponents,  we performed Monte-Carlo simulations on Cr$_{1.38}$Te$_2$ for the 3D-Ising model (see Supplemental information for more details). Figs.~\ref{7}(a), ~\ref{7}(b), and ~\ref{7}(c) show the simulated temperature-dependent magnetization [$M(T)$], susceptibility [$\chi(T)$], and magnetic specific heat [$C_m(T)$] data following the Eqns.~\ref{Eq10},~\ref{Eq11}, and~\ref{Eq12}, respectively. Importantly, the $M(T)$ data exhibit an ascending magnetization trend as the system is cooled and shows a paramagnetic to a ferromagnetic transition at a critical temperature of $T_C\approx$319 K. This critical temperature is very close to the experimental value of 316 K. From the magnetic susceptibility and magnetic specific heat data, depicted in Figs.~\ref{7}(b) and ~\ref{7}(c), the peaks observed at 319 K again remarkably support the critical temperature obtained from the simulated magnetization. Further with the help of simulated magnetic data, we determined the critical exponents $\nu=0.7922\pm$0.024, $\beta=0.4133\pm$ 0.0013, and $\gamma=1.1758\pm$0.0029 (see Supplemental information), consistent with the experimental values. Critical exponents of Cr$_{1.38}$Te$_2$ obtained by various techniques such as MAP, KF, critical isotherm analysis, and Monte-Carlo simulations are tabulated in Tab~\ref{T1}. The critical exponents from various standard models such as the Landau meanfield, 3D Ising, and 3D Heisenberg are also listed in Tab~\ref{T1}.

\begin{equation}
     M = <\frac{1}{N}\sum_i S_i>
    \label{Eq10}
\end{equation}

\begin{equation}
    \chi = \frac{<M^2>-<M>^2}{k_BT}
    \label{Eq11}
\end{equation}
 										
\begin{equation}
    C_m = \frac{<E^2>-<E>^2}{k_BT^2}
    \label{Eq12}
\end{equation}

\begin{figure}[htbp]
    \centering
    \includegraphics[width=0.7\linewidth]{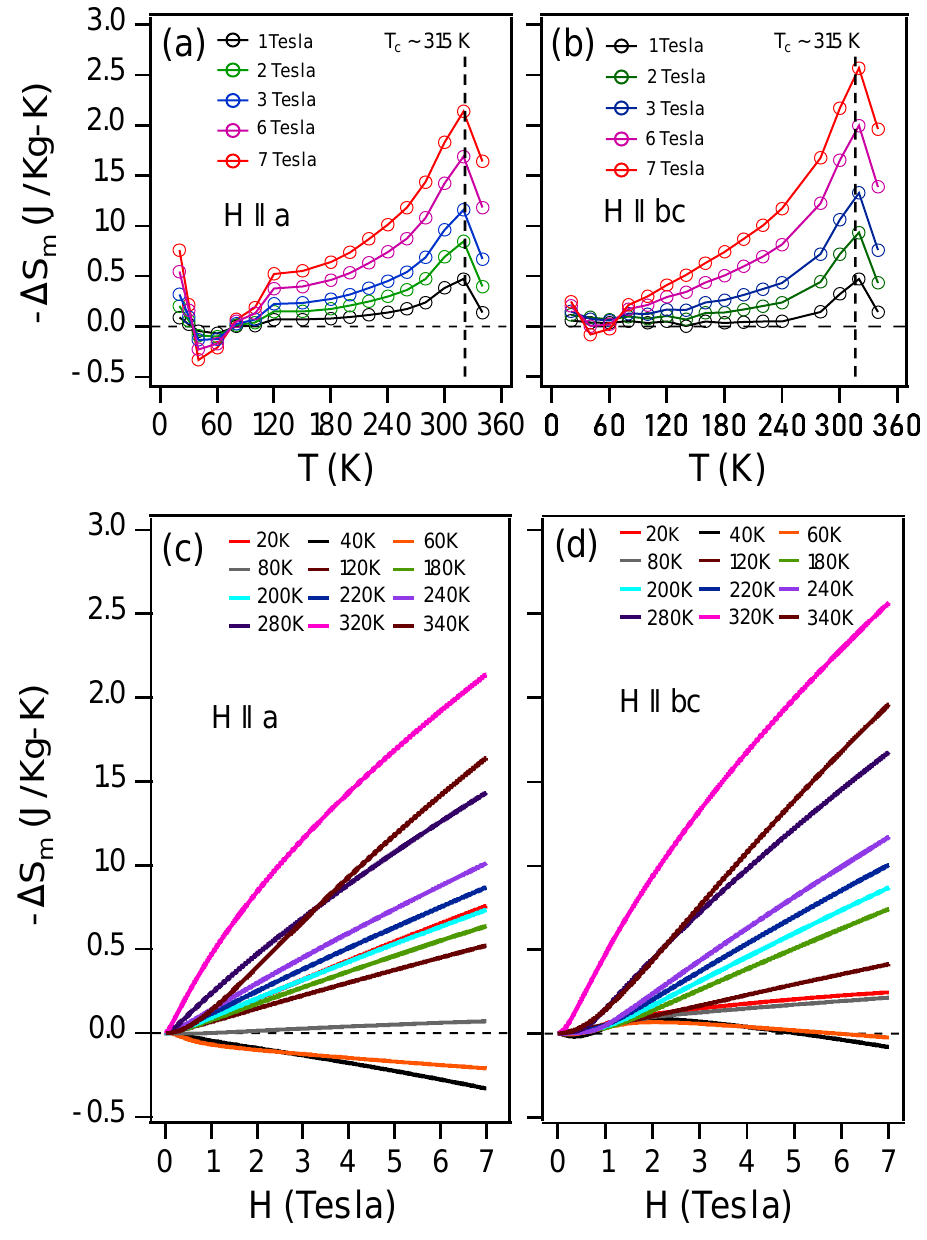}
    \caption{Calculated magnetic entropy change -$\Delta S_m$ as a function of temperature under different magnetic fields for (a) $H\parallel a$ and (b) $H\parallel bc$ .  (c) Magnetic entropy change (-$\Delta S_m$) plotted as a function of field at different sample temperatures for  (c) $H\parallel a$ and (d) $H\parallel bc$.}
    \label{8}
\end{figure}

\subsection{ Magnetocaloric Effect}
With the help of magnetization isotherms [$M(T)$] shown in Figs.~\ref{4}(a) and \ref{4}(b), we investigated the magnetocaloric effect (MCE) around the  $T_C$ in Cr$_{1.38}$Te$_2$ for $H\parallel bc$ and $H\parallel a$ orientations due to potential applications in the magneto-refrigeration technology. Magnetocaloric effect is an intrinsic property of a ferromagnetic system which causes heating or cooling effect, adiabatically,  under the influence of external magnetic fields~\cite{Pecharsky1999}. Thus, a magnetic entropy change -$\Delta S_m (T, H)$ is induced in the presence of magnetic fields which is represented by,
\begin{equation}
\Delta S_m (T, H) =  \int_{o}^{H} (\frac{\partial S}{\partial H})_T dH = \int_{o}^{H} (\frac{\partial M}{\partial T})_H dH
 \label{Eq14}
 \end{equation}
here ($\frac{\partial S}{\partial H})_T$ = ($\frac{\partial M}{\partial T})_H$  follows the Maxwell’s relation. In case of magnetization measured at small discrete fields and temperature intervals, $\Delta S_m (T, H)$ could be written as
\begin{equation}
\Delta S_m (T, H) = \frac{\int_{0}^{H} M(T_{i+1}, H) dH - \int_{0}^{H} M(T_{i}, H) dH }{T_{i+1}-T_i}
 \label{Eq15}
\end{equation}

Figs.~\ref{8}(a) and \ref{8}(b) depict -$\Delta S_m (T, H)$, taken in the temperature range of 5-330 K at various magnetic fields varied up to 7 T, respectively for  $H\parallel a$ and $H\parallel bc$ orientations. The -$\Delta S_m (T, H)$ curves exhibit a maximum (normal MCE) around the PM-FM magnetic transition (T$_C$ $\approx$ 315 K) and a minimum (inverse MCE) around the FM-AFM transition of $T_{N}$ $\approx$ 50 K~\cite{Purwar2023}.  Further, the value of -$\Delta S_m (T, H)$ increases monotonically with field for both orientations as shown in Figs.~\ref{8}(c) and \ref{8}(d), at all the measured temperatures down to 50 K. But below 50 K, -$\Delta S_m (T)$  various non-monotonically with field, which is consistent with -$\Delta S_m (H)$  shown in Figs.~\ref{8}(a) and \ref{8}(b). As discussed in the $structural~properties$ section, the non-monotonic changes of lattice distortion induced by the magneto-volume effect  might be playing the crucial role in the non-monotonic changes of -$\Delta S_m (T)$. Further, under the field of 7 T around $T_C$, the maximum value of - $\Delta S_m (T, H)$ is about 2.08 J kg$^{-1}$ K$^{-1}$ for $H\parallel a$ and is about 2.51 J kg$^{-1}$ K$^{-1}$  for $H\parallel bc$. These -$\Delta S_m (T, H)$ values taken at 7 T are comparable to the values obtained from the other 2D magnetic systems such as Cr$_2$Ge$_2$Te$_6$ (2.64 J kg$^{-1}$ K$^{-1}$)~\cite{Liu2019a} and Cr$_5$Te$_8$ (2.38 J kg$^{-1}$ K$^{-1}$)~\cite{Liu2019} and larger than the values of Fe$_{3-x}$GeTe$_2$ (1.14 J kg$^{-1}$ K$^{-1}$)~\cite{Liu2019b} and CrI$_3$ (1.56J kg$^{-1}$ K$^{-1}$)~\cite{Liu2018}. However,  significantly smaller than the values of  CrBr$_3$ (7.2 J kg$^{-1}$ K$^{-1}$)~\cite{Yu2019} and Cr$_2$Si$_2$Te$_6$ (5.05 J kg$^{-1}$ K$^{-1}$)~\cite{Liu2019a}.


\section{Summary}
In summary, we studied the magnetocrystalline anisotropy, critical behaviour, and magnetocaloric effect in the layered room-temperature monoclinic ferromagnet Cr$_{1.38}$Te$_2$. Our systematic investigation on the structural properties of  Cr$_{1.38}$Te$_2$ as a function of temperature establishes a relation between the magnetic transitions and the crystal lattice. The derived critical exponents $\beta$ = 0.429(5) ($T_C \approx 315.84(2)$ K), $\gamma$ = 1.123(4) ($T_C \approx 315.87(5)$ K) using the KF method, and $\delta$ = 3.662(6) using the CI analysis at T$_C$ = 316 K are self-consistent and obey the rescaling analysis.  The renormalization group (RG) theory suggests that the derived critical exponents of Cr$_{1.38}$Te$_2$ exhibit a three-dimensional (3D) Ising-like long-range exchange interactions [$J(r)$], decaying as $J (r) = r^{-(d+\sigma)}= r^{-4.73}$. The RG theory suggests 3D-Ising type magnetic interactions in this system which is further confirmed by the Monte-Carlo simulations.  Further, the magnetocrystalline anisotropy energy density (K$_u$) is found to be temperature dependent and reaches a maximum (180 kJ/$m^3$) at 110 K. Maximum entropy change -$\Delta S_{m}^{max}$$\approx$2.51 J/ kg-K is found near the $T_C$ for $H\parallel bc$. The density functional theory predicts $b$-axis as the magnetic easy-axis  by analyzing the magnetocrystalline anisotropy energy values for different spin configurations.

\section{Acknowledgements}
This research has made use of the Technical Research Centre (TRC) Instrument Facilities of S. N. Bose National Centre for Basic Sciences, established under the TRC project of the Department of Science and Technology, Govt. of India.
This work is supported by the SERB with grant no. SRG/2022/001102 and “IISER Kolkata Start-up-Grant" Ref.No.IISER-K/DoRD/SUG/BC/2021-22/376 for B. L. C. We acknowledge the support provided by the Kepler Computing facility, maintained by the Department of Physical Sciences, IISER Kolkata, for various computational needs.

\section*{References}
\bibliographystyle{iopart-num}
\bibliography{Cr3Te4}

\end{document}